\documentclass[submitting]{nst}

\usepackage{subfigure,dcolumn}
\usepackage[T2A,T1]{fontenc}
\usepackage[russian,english]{babel}

\begin{document}

\title{Generation of fission yield covariance matrices and its application in uncertainty analysis of decay heat}

\thanks{This work was supported by the National Natural Science Foundation of China (No.12347126) and Presidential Fundation of CAEP (Grand No. YZJJZQ2023022).}

\author{Wendi Chen}
\affiliation{Institute of Applied Physics and Computational Mathematics, Beijing, 100094, China}
\author{Tao Ye}
\email[Corresponding author ]{ye_tao@iapcm.ac.cn}
\affiliation{Institute of Applied Physics and Computational Mathematics, Beijing, 100094, China}
\author{Hairui Guo}
\affiliation{Institute of Applied Physics and Computational Mathematics, Beijing, 100094, China}
\author{Jiahao Chen}
\affiliation{Institute of Applied Physics and Computational Mathematics, Beijing, 100094, China}
\author{Bo Yang}
\affiliation{Institute of Applied Physics and Computational Mathematics, Beijing, 100094, China}
\author{Yangjun Ying}
\affiliation{Institute of Applied Physics and Computational Mathematics, Beijing, 100094, China}

\begin{abstract}
The uncertainties and covariance matrices of fission yield are important in the uncertainty analysis of decay heat. At present, there are no covariance matrixes of fission yield given in the evaluated nuclear data library, although they have provided the uncertainties with good estimates. In this work, the generalized least squares (GLS) updating approach was adopted to evaluate the fission yield covariances with the constraints from basic physical conservation equation and chain yield data, using the nuclear data files from ENDF/B-VIII.0, JENDL-5 and JEFF-3.3. Based on these original and updated data, summation calculation was performed for fission pulse decay heat of thermal neutron-induced fission of $^{235}$U. The uncertainties of decay heat were obtained through generalized perturbation theory, including the uncertainties propagated from fission yield, decay energy, decay constant and branching ratio. The original uncorrelated yield data contributes a $\sim 4 \%$ uncertainty at all times and dominates the decay heat uncertainty at cooling times longer than \SI{100}{s}. With the generated covariance matrixes, the uncertainty of calculated decay heat is strongly reduced and decay energy data makes a major contribution in general. The relative uncertainties at cooling time \SI{0.1}{\second} are $\sim$10$\%$ for ENDF/V-VIII.0 and JEFF-3.3 and $\sim$5$\%$ for JENDL-5 and those at cooling time 10$^{5}$ s are about 1$\%$ for three libraries. The influence of the GLS updating procedure on the contributions of important fission products to decay heat and their sensitive coefficients was also discussed.
\end{abstract}


\maketitle

\section{Introduction}
\label{sec-1}

Nowadays, uncertainty analysis is intensively required in nuclear engineering research. As the major contributor of energy released after the reactor shutdown, the decay heat of fission products is crucial in various aspects of nuclear engineering, such as reactor design, nuclear power plant safety analysis and nuclear fuel waste management \cite{Zhang2024Prediction,Chang2020Best,Li2022Validation}. Significant efforts have been devoted to the accurate estimation and uncertainty analysis of decay heat, which are usually performed in the framework of summation method \cite{Katakura2013Uncertainty,Chiba2017Consistent,Nichols2023Improving,Wang2021Study,Ma2024Decay}.

The summation calculations of decay heat are dependent on the decay and fission yield data, which have been evaluated and compiled in sub-libraries by the evaluated nuclear data libraries, such as ENDF/B \cite{Brown2018ENDF}, JENDL \cite{Iwamoto2023Japanese} and JEFF \cite{Plompen2020Joint}. However, these libraries contain only the best-estimates and uncertainties, without the covariances for fission yield and branching ratio. Especially, the covariances of fission yields play an important role in the uncertainty analysis related to fission products and can strongly reduce the impact of fission yield on the total uncertainty of the system \cite{Katakura2012jendl,Fiorito2015137,Wang2020Lognormal,Lu2023Generation}. 

Several efforts have been made to provide fission yields with complete covariance matrices. A relatively simple procedure was provided by Katakura \cite{Katakura2012jendl,Katakura2013Uncertainty}, in which the chain yields are regarded as the mass distributions of independent fission yields approximately and used as a single constraint. The covariances given by this method only exist between the fission products with the same mass number. Furthermore, constraints from experimental data and physical model were applied to generate the optimized fission yield data and the correlations between all isotopes \cite{Fiorito2016Generation,Tsubakihara2021Evaluation,Matthews2021Stochastically,Lovell2023Calculated,Shu2022fission,Lu2023Generation}. Recently, different kinds of neural networks were adopted to estimate fission yields and their uncertainties \cite{Wang2019Bayesian,Wang2021Optimizing,Qiao2021Bayesian,Xiao2023Bayesian,Song2023Verification,Huo2023Evaluation}, in which the complicated data relationships and the propagation of high order errors could be handled naturally. At present, an international effort coordinated by IAEA is ongoing with the aim of updating and improving the fission yield of actinide nuclides \cite{Vogt2024summary}.

In this work, we focused on the generation of fission yield covariances, in which the constraints from chain yield data and physical conservation equation were considered. Covariance matrices were generated within the generalized least square (GLS) updating approach \cite{Smith1991probability} and a little adjustment was made on the fission yield data simultaneously. Fission pulse decay heat (FPDH) calculations were performed for thermal neutron-induced fission of $^{235}$U, which is denoted by $^{235}$U($n_{\mathrm{th}},f$). The original data in ENDF/B-VIII.0, JENDL-5 and JEFF-3.3 \cite{Brown2018ENDF,Iwamoto2023Japanese,Plompen2020Joint} and the corresponding updated data were used.The uncertainty analysis was carried out with generalized perturbation theory \cite{Jeffery1960Time,Chiba2015Uncertainty}, including the uncertainties propagated from fission yield and decay data.

The paper is organized as follows. The theoretical methodologies of the covariance generation, summation calculation and generalized perturbation theory are described in Sec. \ref{sec-2}. The calculated results and discussions for decay heat and its uncertainties are presented in Sec. \ref{sec-3}. Finally, the conclusion is given in Sec. \ref{sec-4}.

\section{Methodology}
\label{sec-2}

\subsection{Independent fission yield data and covariance generation}
\label{sec-2-1}

Independent fission yield (IFY) $Y_{\mathrm{I}}(A,Z,M)$ is required in decay heat calculations, which represents the probability of a particular nucleus with mass $A$, charge $Z$ and isomeric state $M$ produced directly from one fission, after the emission of prompt neutrons and photons, but before the emission of delayed neutrons. The most used general-purpose evaluated nuclear data libraries: ENDF/B, JENDL and JEFF, provide IFY data with their uncertainties as standard deviation. ENDF/B-VIII.0 takes binary fission into account only, while JENDL-5 and JEFF-3.3 both include the light products of ternary fission in addition. To date, no correlation between fission yields is supplied in these libraries.

The IFY data in these libraries are given by empirical models in general, which are constrained by physical conditions and conservation equations, reflecting the general nature of a fissioning system. These properties should be held in the covariance generation for IFYs, corresponding to the constraint conditions in the GLS updating process. In the present work, the following constraints are employed:

(1) Conservation of mass and charge numbers

For a real fission event, the compound nucleus (CN) would split into two fission fragments generally, named binary fission. These fragments will decay into fission products (FPs) by releasing neutrons and photons. Furthermore, ternary fission may happen with a small probability ($\sim 0.2-0.5 \%$), producing an extra light-charged particle (LCP) such as proton, triton and $\alpha$ particles \cite{Schunck2022Theory}. It should be emphasised that light charged particles were also included in fission products in this paper and the yield data of LCPs and the relatively heavy fission products would be updated together within the GLS updating process.

The conservation laws say that the mass and charge numbers of the compound nucleus must be conserved as follows:
\begin{equation}\label{e-cons-1}
\sum_i{A_iY_{\mathrm{I}}\left( i \right)}=\boldsymbol{A}^t\boldsymbol{Y}_I=A_{\mathrm{CN}}-\bar{\nu}_p ,
\end{equation}

\begin{equation}\label{e-cons-2}
\sum_i{Z_iY_{\mathrm{I}}\left( i \right)}=\boldsymbol{Z}^t\boldsymbol{Y}_I=Z_{\mathrm{CN}} ,
\end{equation}
where $A_{\mathrm{CN}}$ and $Z_{\mathrm{CN}}$ represent the mass and charge numbers of compound nucleus respectively. The mass number, charge number and isomeric state of $i$-th FP are $A_i$, $Z_i$ and $M_i$ respectively. Its independent yield is $Y_{\mathrm{I}}(i)=Y_{\mathrm{I}}(A_i,Z_i,M_i)$. $\boldsymbol{A}$,  $\boldsymbol{Z}$ and $\boldsymbol{Y}_{\mathrm{I}}$ are the vectors containing the mass number, charge number and IFY of each FP. The superscript $t$ denotes the transposition for vector or matrix. $\bar{\nu}_p$ is the average number of prompt neutrons. $A_i$ ranges from 66 to 172 for IFYs in ENDF/B-VIII.0, as only the binary fission is considered in this library. The range of $A_i$ will be expanded for yield data in JENDL-5 and JEFF-3.3 to include light-charged particles. Similar things occur on the charge numbers of fission products $Z_i$.

(2) Normalization of IFYs

The sum of IFYs for FPs except LCPs should be 2 by its definition, that is,
\begin{equation}\label{e-cons-3}
\sum_{i\notin \mathrm{LCP}}{Y_{\mathrm{I}}\left( i \right)}=\boldsymbol{H}_1^t\boldsymbol{Y}_I=2.
\end{equation}
The sensitivity vector $\boldsymbol{H}_1$ is a unit vector for ENDF/B-VIII.0, while it will have some zero elements for LCPs in JENDL-5 and JEFF-3.3.

(3) Normalization of heavier mass yields

For a set of IFY data, there should be a midpoint mass number $A_{\mathrm{mid}}$ such that the yields of FPs on either side of this midpoint should be sum to one when the yield data of LCPs are not counted. As there has been a normalization constraint of IFYs to be 2, this condition can be expressed without repetition as
\begin{equation}\label{e-cons-4}
  \sum_{A_i>A_{\mathrm{mid}}}{Y_{\mathrm{I}}\left( i \right)}=\boldsymbol{H}_{2}^{t} \boldsymbol{Y}_I=1 ,
\end{equation} 
where $\boldsymbol{H}_{2}$ is a sensitivity vector with unit coefficients for products with mass numbers $A$ larger than $A_{\mathrm{mid}}$ and zero elsewhere. In fact, Eq. (\ref{e-cons-4}) is an approximate condition as the mass number is an integer rather than a continuous number. Hence $A_{\mathrm{mid}}$ is set to be $( A_{\mathrm{CN}}-\bar{\nu}_p )/2$ approximately.

(4) Charge symmetry

In the low energy neutron-induced fission, FP has little probability of releasing charged particles in its prompt de-excitation process. Therefore, the charge distribution of IFYs should be symmetrical strictly if only binary fission happens, while this symmetry is slightly broken by ternary fission. Following the procedure of Mills \cite{Mills1995fission}, charge symmetry condition is applied to the charge number pairs $(Z,Z_{\mathrm{CN}}-Z)$ with relatively large yields, that is
\begin{equation}\label{e-cons-5}
  \sum_{Z_{i}=Z}{Y_{\mathrm{I}}\left( i \right)}-\sum_{Z_{i}=Z_{\mathrm{CN}}-Z}{Y_{\mathrm{I}}\left( i \right)}=\boldsymbol{H}_{3}^{t}\boldsymbol{Y}_I=0,
\end{equation}
where $\boldsymbol{H}_3$ is a sensitive vector defined as
\begin{equation}\label{e-H2}
\boldsymbol{H}_3\left( i \right) =
\begin{cases}
	1,Z_i=Z\\
	-1,Z_i=Z_{\mathrm{CN}}-Z\\
	0,\mathrm{else}\\
\end{cases}
\end{equation}
In practical calculations, this constraint is used for the case where charge yields $Y_{\mathrm{I}}(Z)$ and $Y_{\mathrm{I}}(Z_{\mathrm{CN}}-Z)$ are both larger than 1$\%$. $Y_{\mathrm{I}}(Z) =\sum_{Z_i=Z}{Y_{\mathrm{I}}}$.

(5) Chain yield constrain

Chain yield (ChFY), $Y_{\mathrm{Ch}}(A)$, is defined as the sum of cumulative yields of the last stable nuclei with the same mass number $A$ and is much more precise than IFY experimental data. A strong negative correlation between different nuclides will occur if ChFY is used as a constraint in the evaluation of IFY. 

As noted by England and Rider \cite{England1995evaluation}, the chain yields are evaluated after both prompt and delayed neutron emission. Therefore, the pre-delayed-neutron mass yields of IFYs must be corrected to post-delayed-neutron chain yields. Based on the definition, ChFY can be calculated as 
\begin{equation}\label{e-cons-6}
\begin{aligned}
Y_{\mathrm{C}}\left( i \right) &=Y_{\mathrm{I}}\left( i \right) +\sum_j{b_{ij}Y_{\mathrm{C}}\left( j \right)},
\\
Y_{\mathrm{Ch}}\left( A \right) &=\sum_{A_{i}=A}{d_i Y_{\mathrm{C}}\left( i \right)},
\end{aligned}
\end{equation}
where $Y_{\mathrm{C}}$ denotes the cumulative yield. $b_{ij}$ is the branching ratio that $j$-th nucleus decays to the $i$-th one. $d_i=1$ if the $i$-th nucleus is stable and $d_i=0$ elsewhere. Formally, Eq. (\ref{e-cons-6}) can be written in matrix form $\boldsymbol{Y}_{\mathrm{Ch}}=\boldsymbol{D}^t\boldsymbol{Y}_I$ to describe the relation between the chain yields and independent fission yields. $\boldsymbol{D}^t=\boldsymbol{d}^t\left[ \boldsymbol{E}-\boldsymbol{b} \right] ^{-1}$. $\boldsymbol{E}$ is an unit matrix. $\boldsymbol{d}$ and $\boldsymbol{b}$ are the vector and matrix filled by elements $d_i$ and $b_{ij}$ respectively.

Nowadays, there have been thousands of radionuclides in evaluated nuclear data libraries and their half-lives $T_{1/2}$ vary from $\sim$ 10$^{-20}$ to 10$^{20}$ seconds. However, not all radionuclides should be considered in ChFY calculations. Chain yields are close to the mass distribution of independent fission yields $Y_{\mathrm{I}}\left( A \right) =\sum_{A_i=A}{Y_{\mathrm{I}}}$ and their differences are resulted by the crossing-mass-chain decay processes, in which the parent and daughter nuclides are in different mass chains. In the actual measurement of ChFY, the crossing-mass-chain decay pathway will have almost no contribution to the divergence between $Y_{\mathrm{Ch}}(A)$ and $Y_{\mathrm{I}}(A)$ if the half-life of parent nuclide is too long. Hence, it is practical to choose a suitable cut-off value for half-life $T_{\mathrm{cut}}$. The decay process of a radionuclide is taken into account only when its half-life is no larger than $T_{\mathrm{cut}}$. Otherwise, this radionuclide will be regarded as a stable nucleus. It should be emphasised that this half-life cut-off is only valid for ChFY calculations, which means that the influence of the crossing-mass-chain decay processes of radionuclides with $T_{1/2} > T_{\mathrm{cut}}$ on chain yields is ignorable. In fact, these relatively long-lived radionuclides can still influence the time evolution of fission product composition and the true value of cumulative yields.

\begin{figure}[tbp]
    \centering 
    \includegraphics[width=0.45\textwidth]{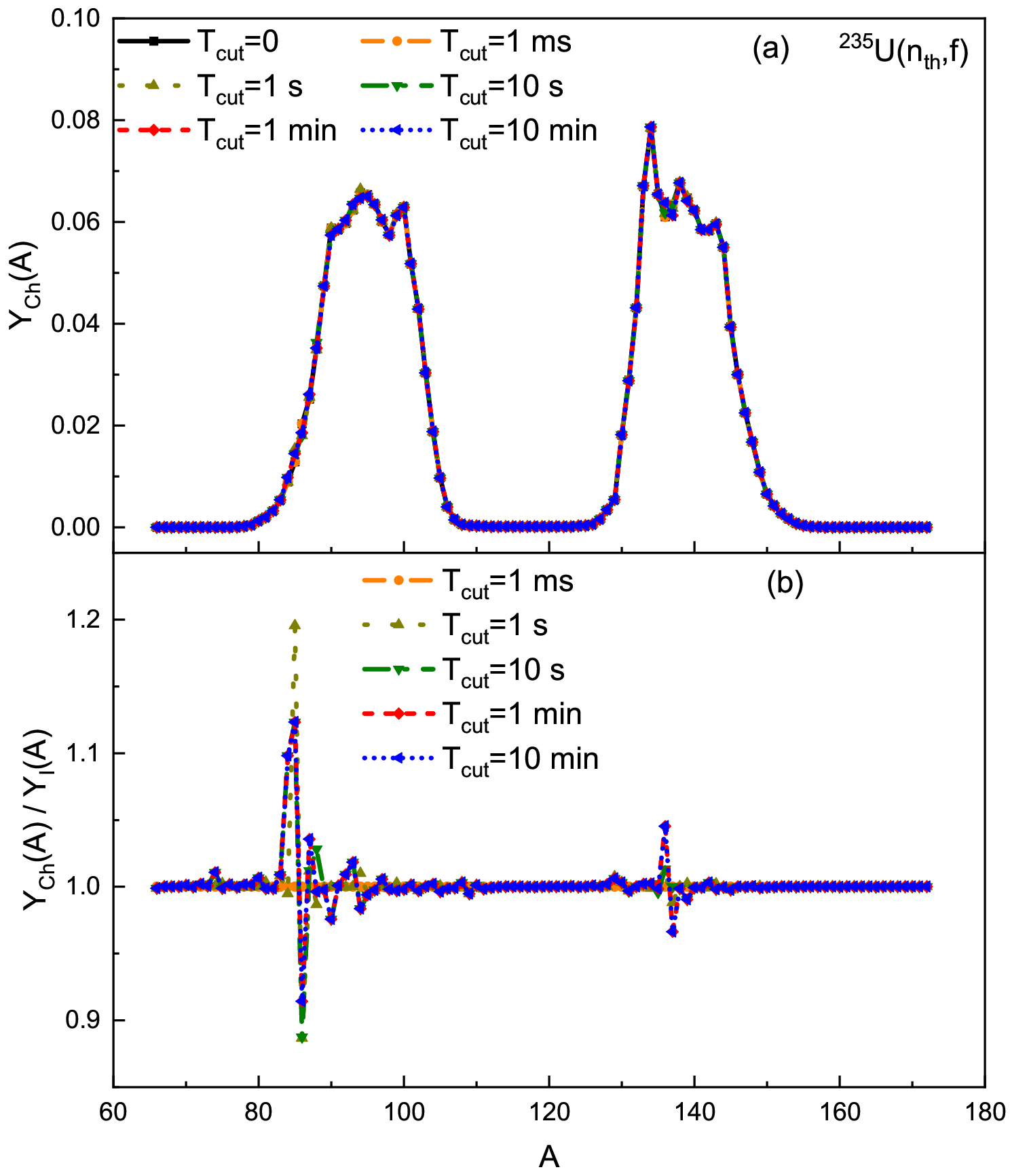}	
    \caption{(color online) (a) Calculated chain yields for thermal neutron-induced fission of $^{235}$U with different half-life cutoff $T_{\mathrm{cut}}$, using the fission yield and decay data in ENDF/B-VIII.0. The result with $T_{\mathrm{cut}}=0$ is equivalent to the mass distribution of IFYs $Y_{\mathrm{I}}(A)$. Solid, dashed, dotted, dash-dotted, short dashed and short dotted lines represent the results calculated with $T_{\mathrm{cut}}$=0, \SI{1}{\ms}, \SI{1}{\second}, \SI{10}{\second}, \SI{1}{\minute} and \SI{10}{\minute} respectively. (b) The ratio of ChFY calculated with non-zero $T_{\mathrm{cut}}$ to $Y_{\mathrm{I}}(A)$. See text for details. } 
    \label{fig-Ych-Tcutoff}
\end{figure}

Fig. \ref{fig-Ych-Tcutoff} shows the calculated ChFYs for thermal neutron-induced fission of $^{235}$U, using the fission yield and decay data in ENDF/B-VIII.0. When $T_{\mathrm{cut}}=0$, all decay processes are eliminated in ChFY calculations and the result is equivalent to the mass distribution of independent fission yields. With the increasement of $T_{\mathrm{cut}}$, the calculated $Y_{\mathrm{Ch}}\left( A \right)$ departs from $Y_{\mathrm{I}}(A)$ gradually and it reaches convergence at $T_{\mathrm{cut}}$=\SI{1}{\minute}. It can be seen that the relative divergences between the converged $Y_{\mathrm{Ch}}(A)$ and $Y_{\mathrm{I}}(A)$ can up to $\sim$ 10$\%$, mainly locating in the mass region $A$=83-97 and 135-140. These differences are contributed primarily by the crossing-mass-chain decay processes of radionuclides with half-lives between \SI{1}{\ms} to \SI{1}{\minute}. The decay processes of nuclides with $T_{1/2}>$ \SI{1}{\minute} have hardly any change in chain yields. Hence $T_{\mathrm{cut}}$=\SI{1}{\minute} is adopted in the following calculations. This cutoff is consistent with the evaluation of England and Rider \cite{England1995evaluation}, in which the half-lives of involved delayed neutron precursors are also no longer than \SI{1}{\minute}.

Generalized least squares method \cite{Smith1991probability,Fiorito2016Generation} is adopted to estimate the covariance matrices of IFYs. For a certain fissioning system, the independent fission yields in the evaluated library are regraded as the prior knowledge of parameters $\boldsymbol{\theta }_a=\boldsymbol{Y}_{\mathrm{I}}$, along with a diagonal prior covariance matrix $\boldsymbol{V}_a$ whose diagonal elements are taken from the corresponding variances. The above constraints can be summarized as an observation equation $\boldsymbol{\eta} \sim \boldsymbol{y}_a=\boldsymbol{S}^t \cdot \boldsymbol{\theta } _a$, where $\boldsymbol{\eta}$ is a vector and denotes one of the above constraints. $\boldsymbol{S}$ is the design matrix. In the GLS approach, the prior parameters and covariance can be updated as
\begin{equation}\label{e-GLS}
\begin{aligned}
\boldsymbol{\theta }_{\mathrm{upd}} &=\boldsymbol{\theta }_a+\boldsymbol{V}_a\boldsymbol{S}^t [ \boldsymbol{S}^t\boldsymbol{V}_a\boldsymbol{S}+\boldsymbol{V} ] ^{-1}\left( \boldsymbol{\eta }-\boldsymbol{y}_a \right), 
\\
\boldsymbol{V}_{\mathrm{upd}} &=\boldsymbol{V}_a-\boldsymbol{V}_a\boldsymbol{S}^t [ \boldsymbol{S}^t\boldsymbol{V}_a\boldsymbol{S}+\boldsymbol{V}  ] ^{-1}\boldsymbol{SV}_a ,
\end{aligned}
\end{equation}
where $\boldsymbol{\theta }_{\mathrm{upd}}$ and $\boldsymbol{V}_{\mathrm{upd}}$ are the posterior best-estimates and covariance for IFYs respectively. $\boldsymbol{V}$ is the covariance matrix of $\boldsymbol{\eta}$. 

In the present work, $\boldsymbol{V}$ is also a diagonal matrix with diagonal elements taken from the variances of $\boldsymbol{\eta}$. The uncertainties of $\bar{\nu}_p$ and chain yield are used directly for the mass conservation and chain yield constraints, and fairly small errors are set for other constraints: 10$^{-6}$ for normalization, 0.5$\%$ for charge symmetry as in the order of the probability of ternary fission and 0.01$\%$ for the rest. It should be stressed that this paper focuses on the fission yield covariance generation and its application in decay heat uncertainty analysis, rather than fission yield data evaluation. Hence, $\bar{\nu}_p$ and its uncertainty in each library are used in the GLS updating procedure for IFYs in the corresponding library. Similarly, the chain yield data of England and Rider \cite{England1995evaluation} are used as constraints for IFYs in ENDF/B-VIII.0 and JENDL-5 and those of Nichols et al. \cite{Nichols2008handbook} are used for IFYs in JEFF-3.3, as indicted in Refs. \cite{Brown2018ENDF, Tsubakihara2021Evaluation, Mills2017New}. The uncertainties of branching ratios were not taken into account in the chain yield constraint.

\begin{table*}[htbp]
\centering
\footnotesize
\begin{tabular*}{\linewidth}{l c c c c c c} 
 \hline
 $^{235}$U($n_{\mathrm{th}},f$) & \multicolumn{2}{c}{ENDF/B-VIII.0} & \multicolumn{2}{c}{JENDL-5} & \multicolumn{2}{c}{JEFF-3.3} \\         
                   & ORI & UPD & ORI & UPD & ORI & UPD  \\
 \hline
 $A_{\mathrm{CN}}-\bar{\nu}_p-\sum{AY_{\mathrm{I}}} $ & -1.42E-2 $\pm$ 3.79 & 3.20E-3 $\pm$ 7.48E-3 & -6.40E-3 $\pm$ 3.01 & -3.00E-3 $\pm$ 6.13E-3 & 1.00E-2 $\pm$ 4.18 & 3.20E-3 $\pm$ 1.14E-2 \\
 $Z_{\mathrm{CN}}-\sum{ZY_{\mathrm{I}}} $ & -5.31E-2 $\pm$ 1.51 & 0 $\pm$ 9.97E-5 & -5.45E-2 $\pm$ 1.18 & 0 $\pm$ 1.00E-4 & 1.60E-4 $\pm$ 1.64 & 0 $\pm$ 1.00E-4 \\
 $2-\sum_{i\notin \mathrm{LCP}}{Y_{\mathrm{I}}} $ & 0 $\pm$ 3.46E-2 & 0 $\pm$ 9.85E-7 & 0 $\pm$ 2.55E-2 & 0 $\pm$ 1.00E-6 & 0 $\pm$ 3.52E-2 & 0 $\pm$ 9.94E-7 \\
 $1- \sum_{A_i > A_{\mathrm{mid}}}{Y_{\mathrm{I}}} $ & -7.40E-5 $\pm$ 1.72E-2 & 6.05E-5 $\pm$ 9.55E-5  & 1.32E-4 $\pm$ 1.76E-2 & 4.84E-5 $\pm$ 9.41E-5 & 0 $\pm$ 2.48E-2 & 0 $\pm$ 9.85E-5\\
 $\chi ^2$ for charge symmetry  & 2.23 & 1.50 & 10.36 & 0.80 & 0 & 0.01 \\
 $\chi ^2$ for chain yield  & 8.48 & 3.23 & 9.09 & 2.52 & 9.78 & 0.03 \\
 \hline 
\end{tabular*}
\caption{Impact of GLS updating procedure on the constraints for $^{235}$U($n_{\mathrm{th}},f$). ORI and UPD denote the results before and after the GLS updating process. The first four quantities are residual errors for the observation equation Eq. (\ref{e-cons-1}), (\ref{e-cons-2}), (\ref{e-cons-3}) and (\ref{e-cons-4}) respectively. The data is given in the form $x \pm \delta x$, representing the calculated residual error and its uncertainty respectively. $\chi ^2$ is given for charge symmetry and chain yield constraints. See text for details.}
\label{table-GLS}
\end{table*}

Table \ref{table-GLS} shows the impacts of GLS updating process on each constraint for thermal neutron-induced fission of $^{235}$U. The residual errors $x=\boldsymbol{\eta }-\boldsymbol{S}^t\cdot \boldsymbol{\theta }$ and their uncertainties $\delta x=\sqrt{\boldsymbol{S}^t\cdot \boldsymbol{V}_{\theta}\cdot \boldsymbol{S}}$ are calculated for the observation equations Eq. (\ref{e-cons-1}), (\ref{e-cons-2}), (\ref{e-cons-3}) and (\ref{e-cons-4}), where $\boldsymbol{V}_{\theta}$ is the covariance matrix for $\boldsymbol{\theta }$. The most significant effect is the reduction on the uncertainties of residual errors by orders of magnitude, especially for the mass and charge conservation constraints. The residual errors also decrease after GLS updating process in most cases. 

\begin{figure}[tbp]
    \centering 
    \includegraphics[width=0.45\textwidth]{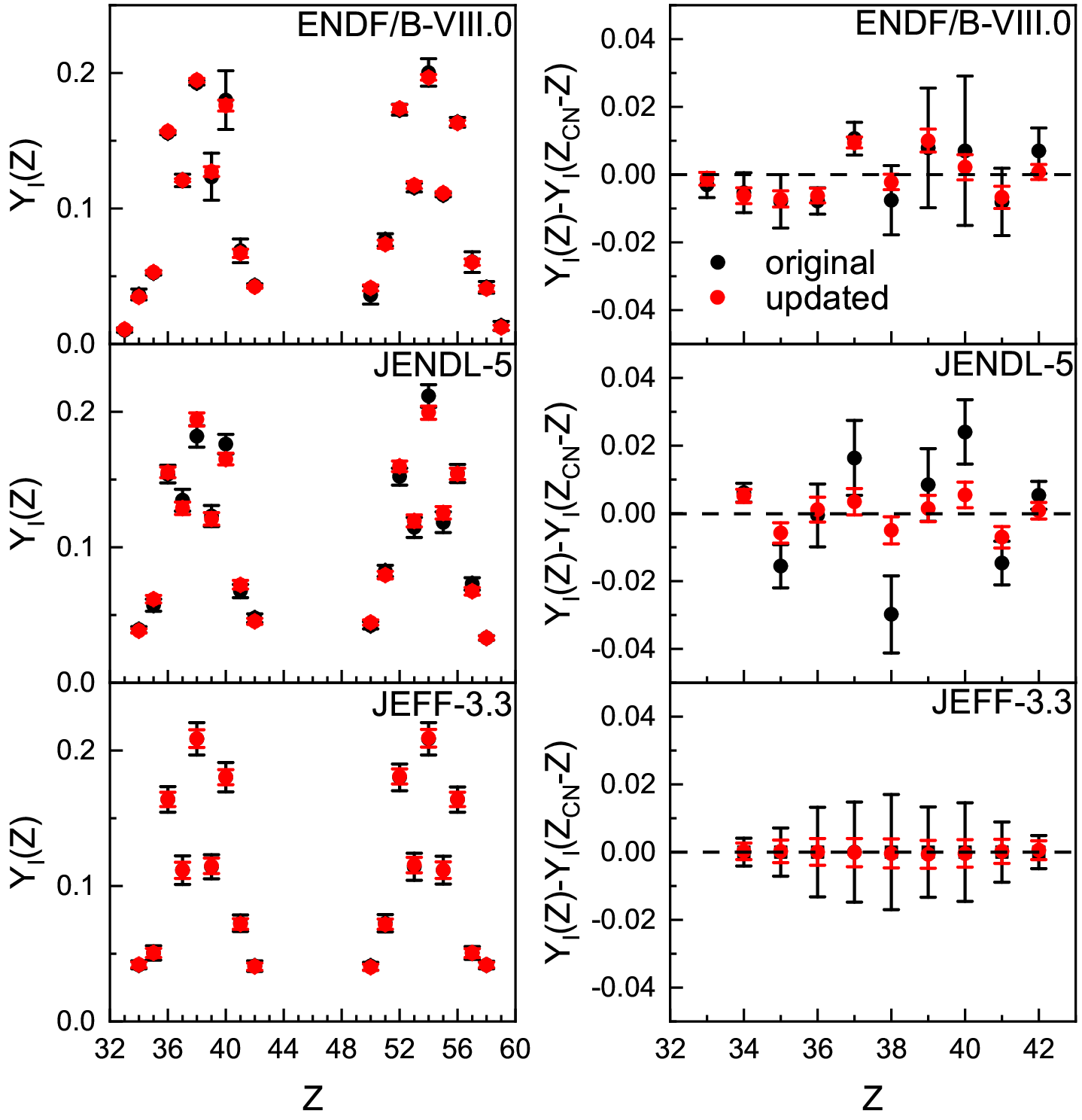}	
    \caption{(color online) (left panel) Comparison of original and updated charge yields $Y_{\mathrm{I}}(Z)$ for $^{235}$U($n_{\mathrm{th}},f$). (Right panel) Comparison of $Y_{\mathrm{I}}(Z)-Y_{\mathrm{I}}(Z_{\mathrm{CN}}-Z)$ calculated with original and updated yield data for $^{235}$U($n_{\mathrm{th}},f$). The black and red circles represent the results calculated with original and updated IFYs. The results are plotted only when $Y_{\mathrm{I}}(Z)$ and $Y_{\mathrm{I}}(Z_{\mathrm{CN}}-Z)$ are both larger than 1$\%$. See text for details.} 
    \label{fig-YZ-235U}
\end{figure}

\begin{figure}[tbp]
    \centering 
    \includegraphics[width=0.45\textwidth]{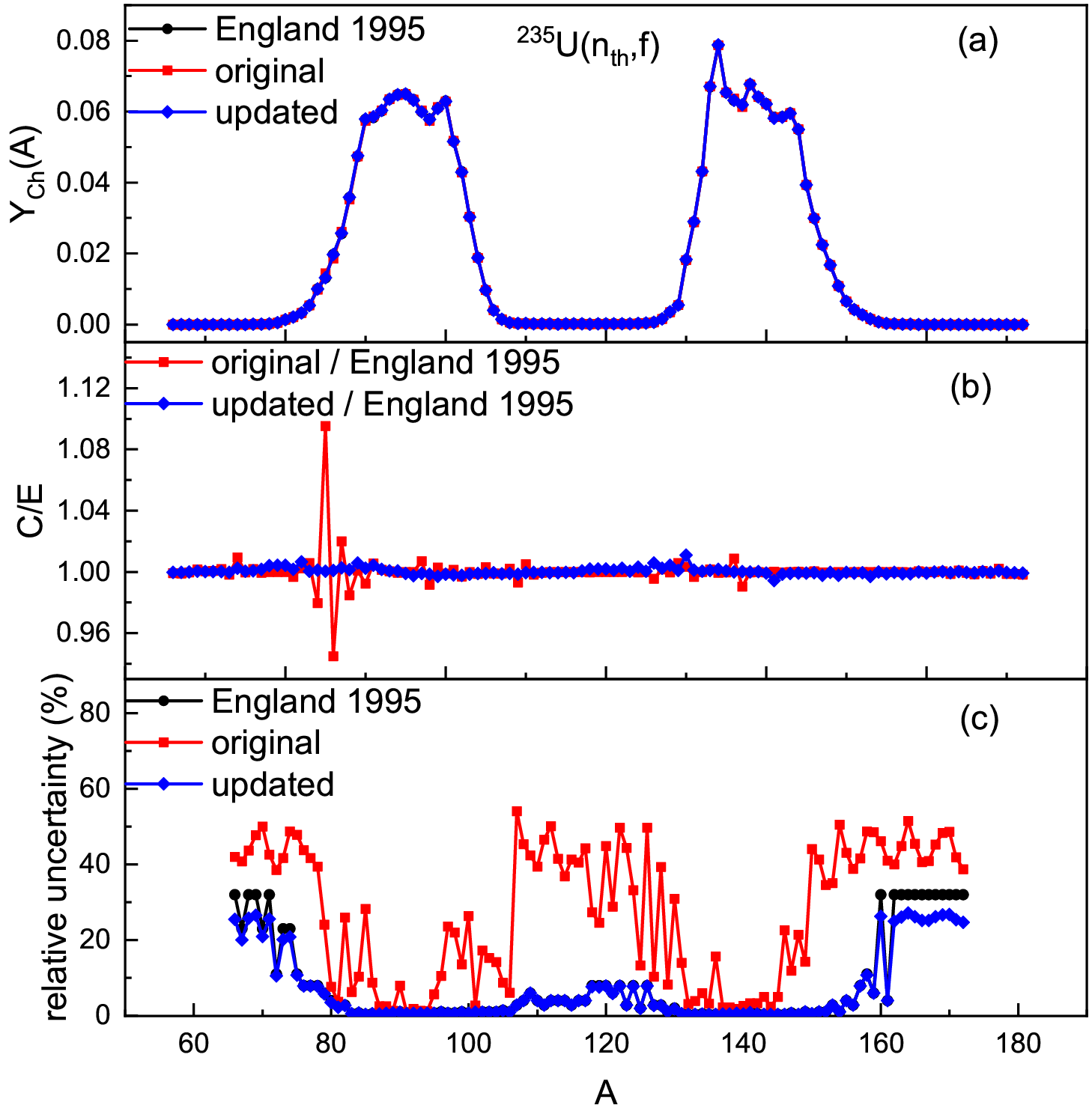}	
    \caption{(color online) Impact of GLS updating procedure on chain yield for $^{235}$U($n_{\mathrm{th}},f$) based on the data file from ENDF/B-VIII.0. Evaluation data are taken from England and Rider \cite{England1995evaluation}. (a) Comparison of original and updated chain yields as well as the evaluation values. (b) Ratio of calculated results to evaluation values (C/E). (c) Comparison of relative uncertainties.} 
    \label{fig-Ych-235U}
\end{figure}

For the charge symmetry and chain yield constraints, the quantity $\chi ^2$ is adopted to assess the improvement quantitatively, which is defined as
\begin{equation}\label{e-GLS-chi2}
\chi ^2=\frac{1}{N} [ \boldsymbol{\eta }-\boldsymbol{S}^t\cdot \boldsymbol{\theta } ]^t \boldsymbol{V}^{-1} [ \boldsymbol{\eta }-\boldsymbol{S}^t\cdot \boldsymbol{\theta } ], 
\end{equation}
where $N$ is the length of vector $\boldsymbol{\eta }$. Strong reduction on $\chi ^2$ is observed in most cases except the charge symmetry constraint for IFY data in JEFF-3.3. It is because the independent yields in JEFF-3.3 \cite{Mills2017New} have been adjusted to ensure the equality of complement charge yields with higher precision than the present work. However, it does not affect the effectiveness of GLS updating procedure. Fig. \ref{fig-YZ-235U} shows the original and updated complementary charge yields as well as $Y_{\mathrm{I}}(Z)-Y_{\mathrm{I}}(Z_{\mathrm{CN}}-Z)$ for $^{235}$U($n_{\mathrm{th}},f$). For the fission yield data in ENDF/B-VIII.0 and JENDL-5, GLS updating procedure slightly adjusts IFYs to reduce the deviations between $Y_{\mathrm{I}}(Z)$ and $Y_{\mathrm{I}}(Z_{\mathrm{CN}}-Z)$ and generates covariance matrix to decrease the uncertainties of $Y_{\mathrm{I}}(Z)$ and $Y_{\mathrm{I}}(Z)-Y_{\mathrm{I}}(Z_{\mathrm{CN}}-Z)$. For JEFF-3.3, there is a noticeable reduction on the uncertainties of $Y_{\mathrm{I}}(Z)$ and $Y_{\mathrm{I}}(Z)-Y_{\mathrm{I}}(Z_{\mathrm{CN}}-Z)$, while $Y_{\mathrm{I}}(Z)$ themselves remain almost unchanged. Similarly, the improvement of agreement with chain yield data also occurs along with the reduction of uncertainties as presented in Fig. \ref{fig-Ych-235U}. A noteworthy thing is that the uncertainties of calculated chain yields are reduced to no more than those of evaluation values in all mass regions after GLS updating process.

\begin{figure}[tbp]
    \centering 
    \includegraphics[width=0.45\textwidth]{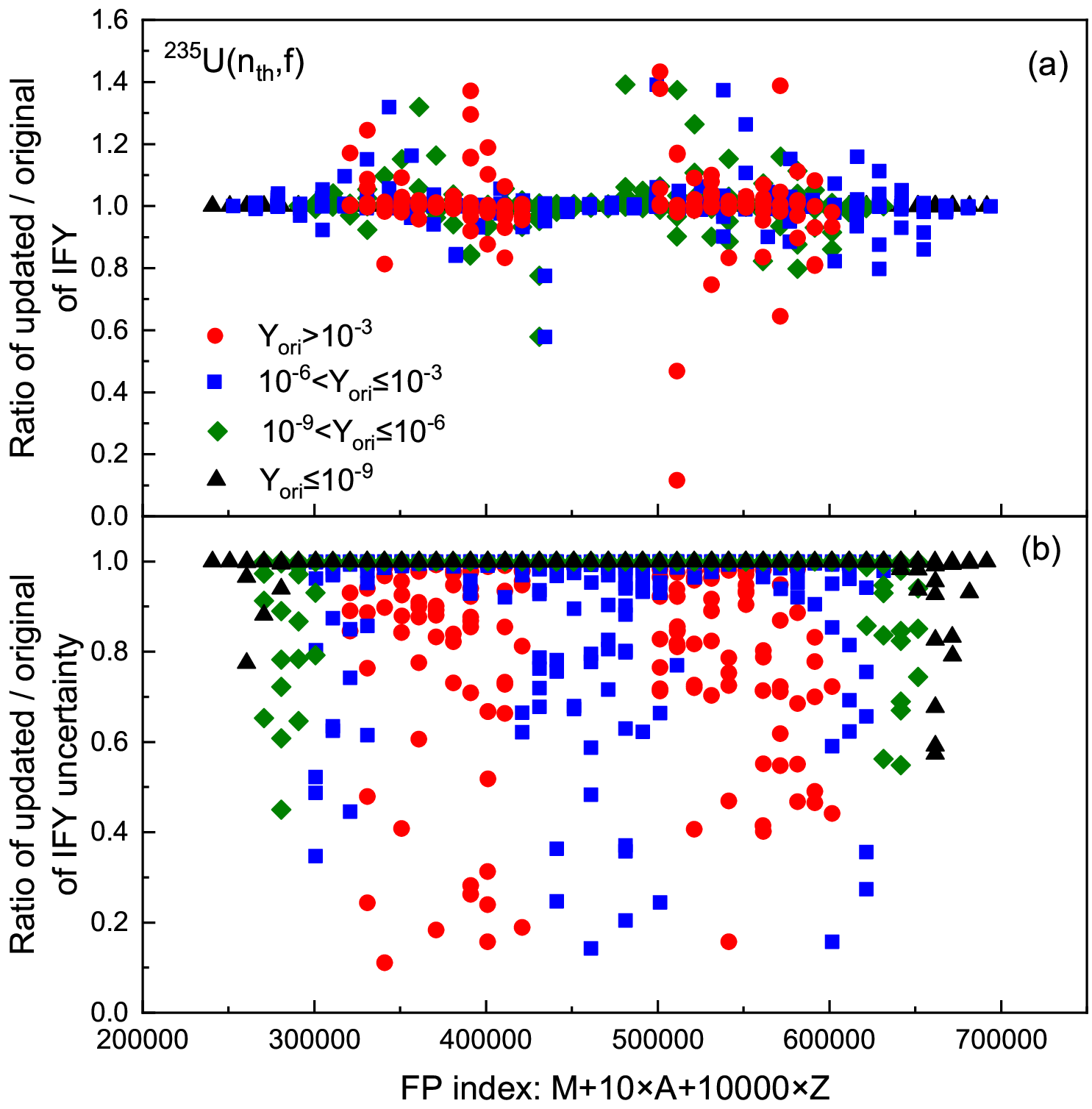}	
    \caption{(color online) Ratio of updated to original data for (a) IFYs $Y_{\mathrm{I}}(A,Z,M)$ and (b) their uncertainties for $^{235}$U($n_{\mathrm{th}},f$) based on the data file from ENDF/B-VIII.0. The values are denoted by circles, squares, rhombuses and triangles when the original IFYs ($Y_{\mathrm{ori}}$) are larger than 10$^{-3}$, in the regions 10$^{-6}$-10$^{-3}$,  10$^{-9}$-10$^{-6}$ and smaller than 10$^{-9}$ respectively. See text for details. } 
    \label{fig-compare-IFY-235U}
\end{figure}

As shown in Eq. (\ref{e-GLS}), GLS updating procedure not only generates the correlation of fission yields but also adjusts the yield data. Comparison of original IFYs in ENDF/B-VIII.0 with updated IFYs are presented in Fig. \ref{fig-compare-IFY-235U} for thermal neutron-induced fission of $^{235}$U. The adjustments are visible only when IFYs have high sensitivity to the constraint system. Strong reductions occur on variances as they are removed from the diagonal part and reintroduced as correlations between IFYs.

\begin{figure}[!htb]
\centering
\subfigure{\includegraphics[width=0.45\textwidth]{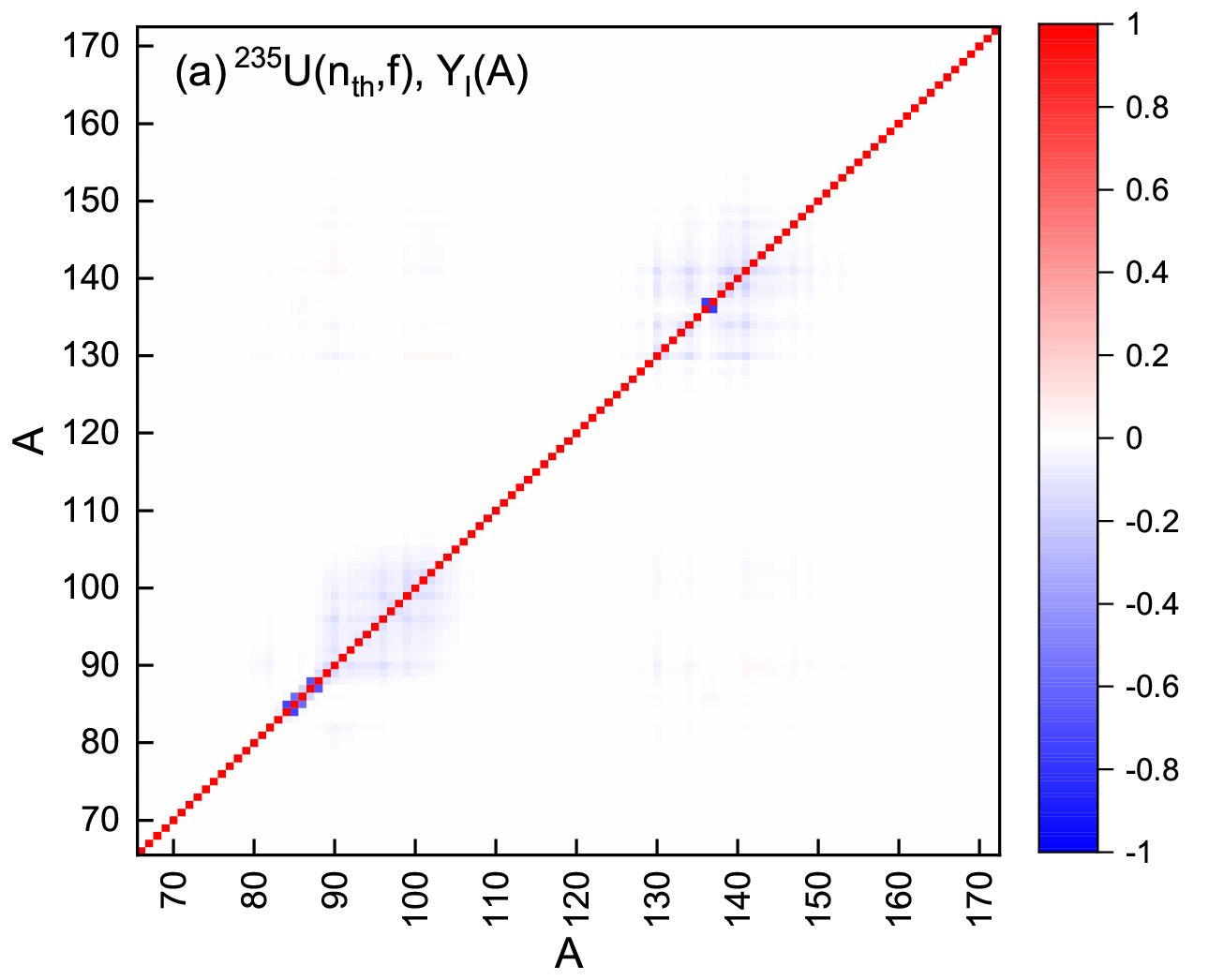}}
\subfigure{\includegraphics[width=0.45\textwidth]{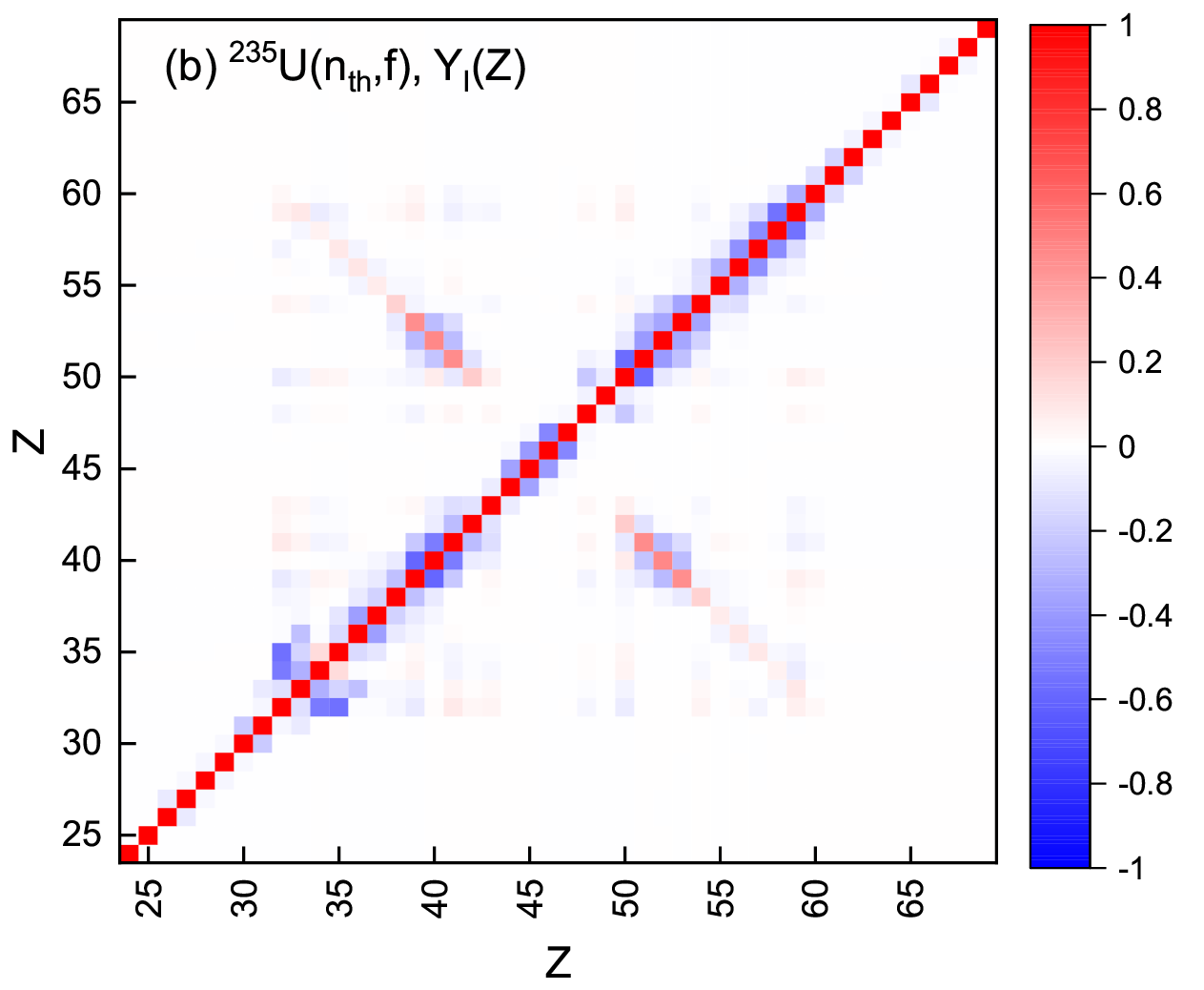}}
\caption{(color online) Correlation matrices for (a) $Y_{\mathrm{I}}(A)$ and (b) $Y_{\mathrm{I}}(Z)$ of updated IFYs based on the data file from ENDF/B-VIII.0 for $^{235}$U($n_{\mathrm{th}},f$).}
\label{fig-COR}
\end{figure}

Fig. \ref{fig-COR} presents the correlation matrices for $Y_{\mathrm{I}}(A)$ and $Y_{\mathrm{I}}(Z)$ of updated IFYs based on the data file from ENDF/B-VIII.0. The correlation matrix $\bar{\boldsymbol{V}}$ is connected with covariance matrix $\boldsymbol{V}$ as
\begin{equation}\label{e-COR}
\bar{\boldsymbol{V}}_{ij} =\frac{\boldsymbol{V}_{ij}}{\sqrt{\boldsymbol{V}_{ii} \boldsymbol{V}_{jj}}}.
\end{equation}
It can be seen that there are mainly negative correlations between $Y_{\mathrm{I}}(A)$ with neighbouring mass numbers and the correlation becomes insignificant with the increase of the difference in mass. Relatively strong negative correlations are observed in the mass regions $A$=84-85-86, 87-88 and 136-137, which are all located in the regions that the converged $Y_{\mathrm{Ch}}$ differs from $Y_{I}(A)$ observably as shown in Fig. \ref{fig-Ych-Tcutoff}(b). It could be inferred that these relatively strong negative correlations result from the inclusion of crossing-mass-chain decay processes in chain yield constraint. Different from mass yields, there are both positive and negative off-diagonal elements in the correlation matrix for $Y_{\mathrm{I}}(Z)$, revealing a more complicated dependence structure in charge yields.

\subsection{Summation calculation and generalized perturbation theory}
\label{sec-2-2}

Fission pulse decay heat is the heat generated by radioactive decay after an instantaneous burst of fission which happens at time $t$=0. FPDH can be calculated within the summation method as
\begin{equation}\label{e-FPDH}
H\left( t \right) =\sum_i{H_i}=\sum_i{\lambda _i\bar{E}_iN_i\left( t \right)},
\end{equation} 
where $\lambda _i$, $\bar{E}_i$ and $N_i$ are the decay constant, average released energy per decay and number of $i$-th fission product. The released decay energy $\bar{E}$ consists of contributions from the average energy released as the kinetic energy of light-particle transitions $\bar{E}_{\mathrm{LP}}$ (most frequently as $\beta$ emissions), electromagnetic radiation $\bar{E}_{\mathrm{EM}}$ and heavy particles $\bar{E}_{\mathrm{HP}}$. The last term has little contribution so only the light particle and electromagnetic decay heat ($H_{\mathrm{LP}}$ and $H_{\mathrm{EM}}$) calculations were performed in this work.

The time evolution of $N(t)$ follows Bateman equation \cite{Bateman1910solution} as
\begin{equation}\label{e-TD-FP}
\frac{dN_i}{dt}=-\lambda _iN_i+\sum_j{\lambda _jb_{ij}N_j}.
\end{equation}
IFYs are used as the initial condition $N_i(t=0)$. This ordinary differential equation was solved numerically by the algorithm of Austin Ladshaw et al. \cite{Ladshaw2020Algorithms}.

The uncertainty of FPDH was carried out with generalized perturbation theory \cite{Jeffery1960Time,Chiba2015Uncertainty}. The variance of decay heat $V_H$ can be calculated with the Sandwich formula as
\begin{equation}\label{e-DDH}
  V_H=\boldsymbol{S}_{H}^{t}\boldsymbol{V}_{\sigma}\boldsymbol{S}_H,
\end{equation}
where $\boldsymbol{S}_H$ is the sensitivity vector of decay heat with respect to various nuclear data. Sensitivities of decay heat to decay constant and decay energy can be expressed analytically as
\begin{equation}\label{e-sensi}
\begin{aligned}
S_H ( \lambda _i ) &=\frac{\partial H}{\partial \lambda _i}=\bar{E}N_i(t),
\\
S_H ( \bar{E}_i ) &=\frac{\partial H}{\partial \bar{E}_i}=\lambda _iN_i(t),
\end{aligned}
\end{equation}
while those to branching ratio and IFY are calculated numerically.

$\boldsymbol{V}_{\sigma}$ is a covariance matrix of nuclear data. The covariance of fission yield has been generated in Sec. \ref{sec-2-1}. $\boldsymbol{V}_{\sigma}$ of branching ratio is meaningful for different decay pathways of a single nuclide because of the normalization condition $\sum_j{b_{ji}}=1$. For $i$-th nucleus, its covariance matrix for branching ratios is given as
\begin{equation}\label{e-cov-BR}
\begin{aligned}
\boldsymbol{V}_{jj}^{(i)} &=\sigma _{ji}^{2}\left( 1-\frac{\sigma _{ji}^{2}}{\sum_m{\sigma _{mi}^{2}}} \right), 
\\
\boldsymbol{V}_{jk}^{(i)} &=-\frac{\sigma _{ji}^{2}\sigma _{ki}^{2}}{\sum_m{\sigma _{mi}^{2}}}, k \neq j,
\end{aligned}
\end{equation}
where $\sigma _{ji}$ is the uncertainty of branching ratio $b_{ji}$. The derivation of this expression is given in Appendix \ref{App-1}. The uncertainty of branching ratio is set to be zero for the radionuclide with only one decay pathway. $\boldsymbol{V}_{\sigma}$ of decay constant and decay energy are both diagonal matrices with diagonal elements taken from their variances. If the uncertainties of $\lambda$, $\bar{E}_{\mathrm{LP}}$ and $\bar{E}_{\mathrm{EM}}$ are not given in the evaluated library, their uncertainties are assumed to be 100$\%$.

\section{Results and discussions}
\label{sec-3}

For each evaluated nuclear data library, two sets of data were used in FPDH calculations:

Set-A: original fission yield data with decay data.

Set-B: GLS updated fission yield data with decay data.

No adjustment was made for decay data in the present work.

\subsection{FPDH results}
\label{sec-3-1}

\begin{figure}[tbp]
    \centering 
    \includegraphics[width=0.45\textwidth]{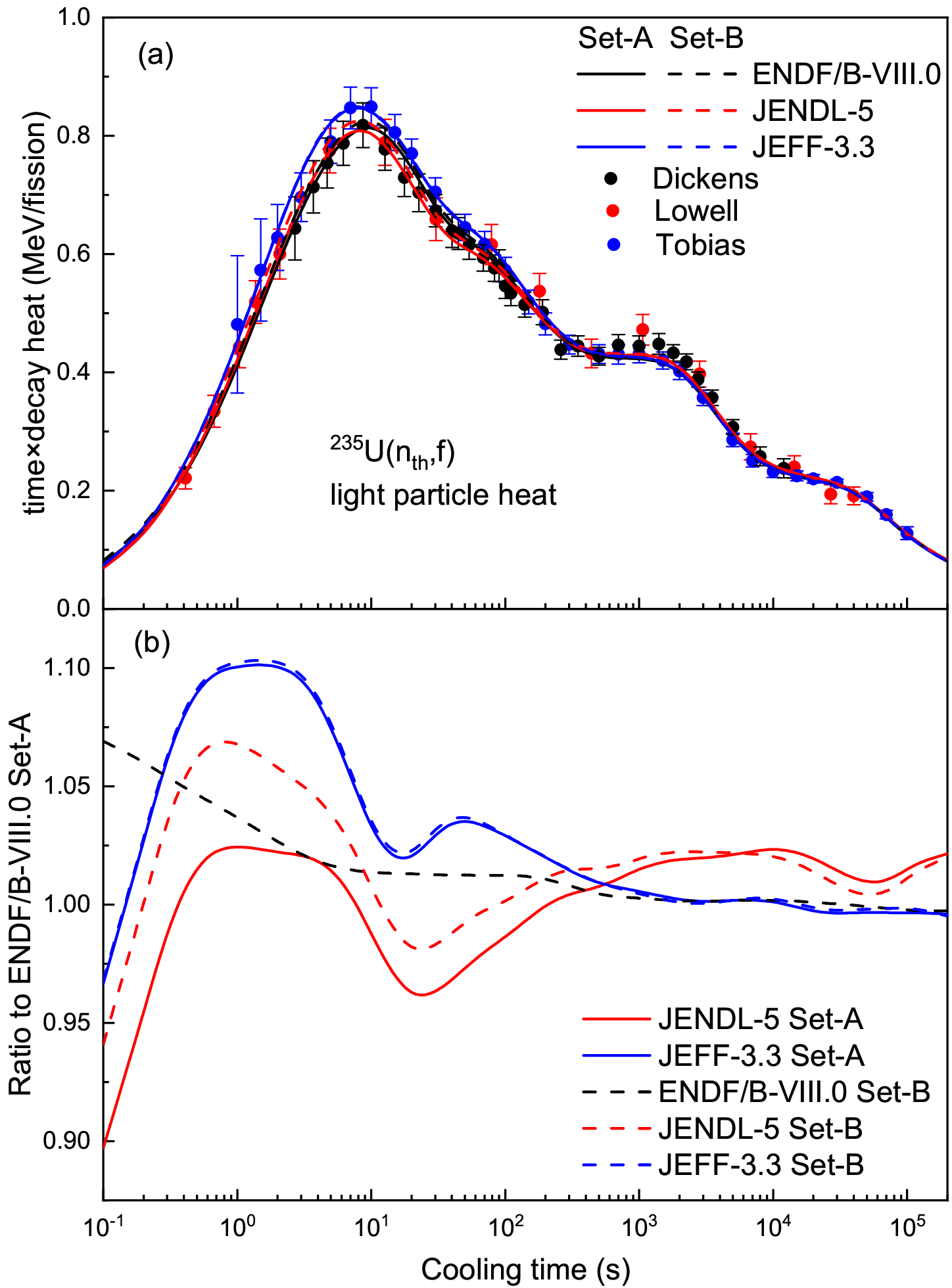}	
    \caption{(color online) (a) Calculated light particle decay heat for $^{235}$U($n_{\mathrm{th}},f$) with Set-A and Set-B. Experimental data are taken from the CoNDERC database \cite{CoNDERC}. The solid and dashed lines represent the calculated results with Set-A and Set-B respectively. The calculated results based on ENDF/B-VIII.0, JENDL-5, JEFF-3.3 are denoted by black, red and blue lines respectively. (b) Ratio of calculated results to that obtained with ENDF/B-VIII.0 Set-A.} 
    \label{fig-HLP-235U}
\end{figure}

\begin{figure}[htbp]
    \centering 
    \includegraphics[width=0.45\textwidth]{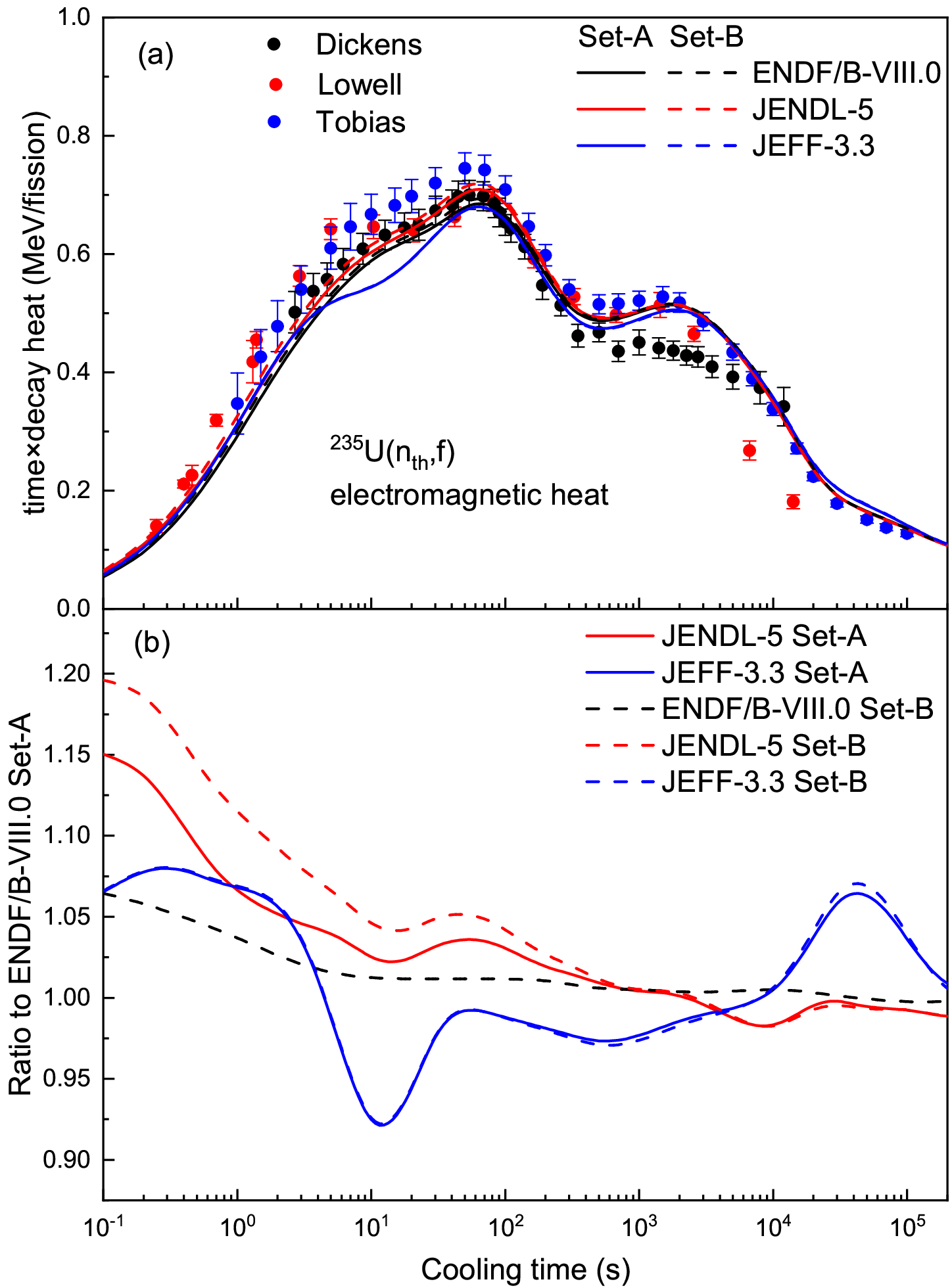}	
    \caption{(color online) Same as Fig. \ref{fig-HLP-235U} but for electromagnetic decay heat. Experimental data are taken from the CoNDERC database \cite{CoNDERC}.} 
    \label{fig-HEM-235U}
\end{figure}

The light particle and electromagnetic components of FPDH for thermal neutron-induced fission of $^{235}$U are presented in Fig. \ref{fig-HLP-235U} and \ref{fig-HEM-235U} respectively, which were calculated with Set-A and Set-B from ENDF/B-VIII.0, JENDL-5, JEFF-3.3. Experimental data are taken from the IAEA CoNDERC database \cite{CoNDERC}. The resulting decay heat curves are in good agreement at cooling times above \SI{1000}{\second}, except that JEFF-3.3 provides a relatively larger electromagnetic decay heat during 10$^{4}$-10$^{5}$ \si{\second}. The discrepancies between calculated results mainly occur within \SI{1000}{\second} and become significant at cooling times shorter than \SI{1}{\second}. In general, all calculated results at cooling times below 1000 s are in the error range of experimental data, except the underestimation of electromagnetic decay heat for JEFF-3.3 from 5 to 50 \si{\second}.

It can be observed for both light particle and electromagnetic decay heats that the updated IFYs lead to an enhancement up to $\sim 5 \%$ at cooling times shorter than 1 s and increase the peaks of curves about 1-2$\%$ for ENDF/B-VIII.0 and JENDL-5, while the divergences between calculated results with Set-A and Set-B are small for JEFF-3.3. These differences reveal that the update on IFYs themselves is moderate for ENDF/B-VIII.0 and JENDL-5 but negligible for JEFF-3.3.

\subsection{Uncertainties of FPDH}
\label{sec-3-2}

The uncertainties of FPDH are propagated from those of fission yield, decay energy, decay constant and branching ratio as formulated in Sec. \ref{sec-2-2}. Computations were carried out with both Set-A and Set-B. In order to perform uncertainty analysis with Set-A data, a diagonal covariance matrix is given for fission yield data, in which the diagonal elements are taken from their variances.

\begin{figure}[!tb]
    \centering 
    \includegraphics[width=0.45\textwidth]{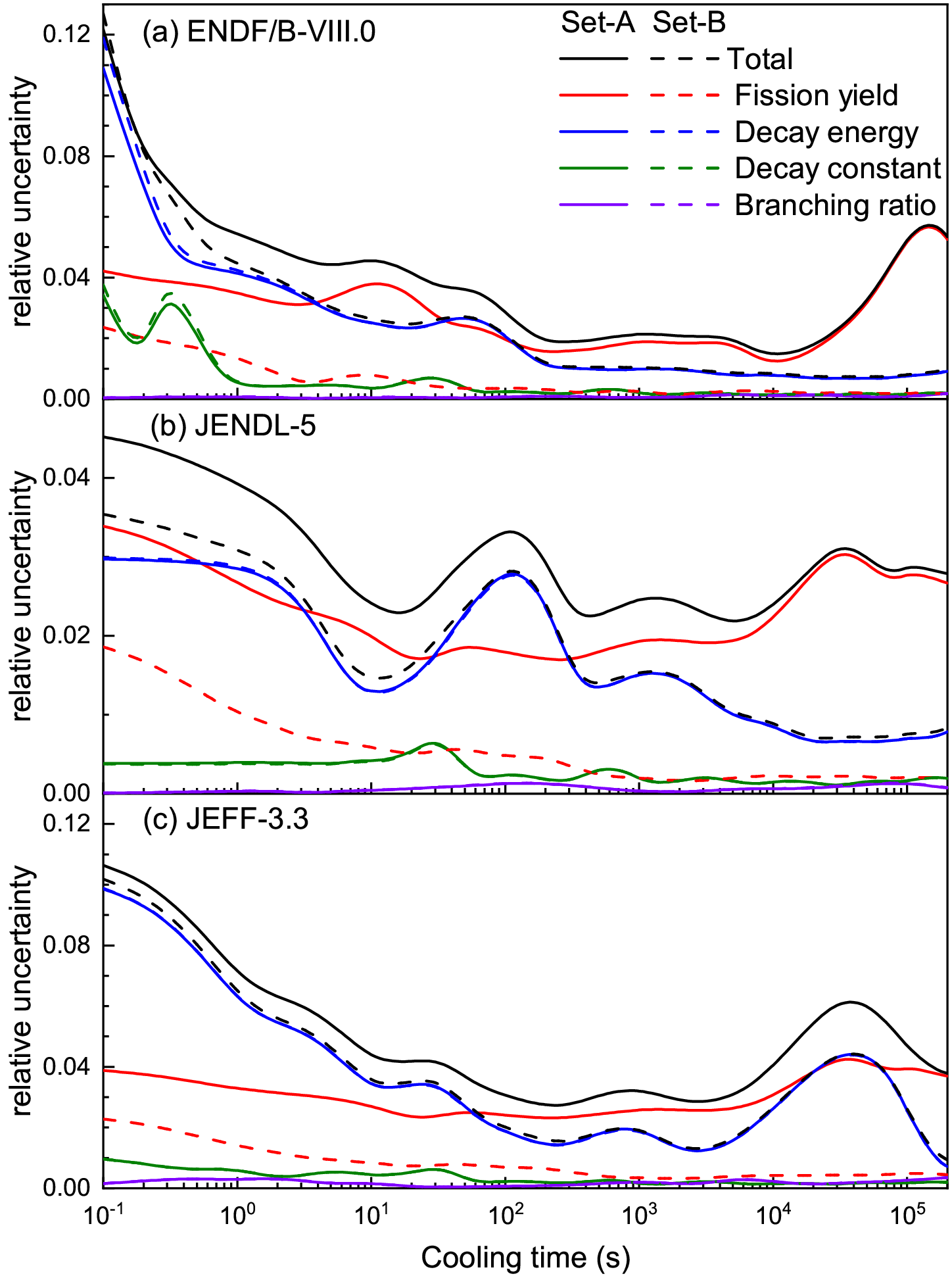}	
    \caption{(color online) Relative uncertainties of light particle decay heat for $^{235}$U($n_{\mathrm{th}},f$). The black lines denote the total uncertainty. The red, blue, green and purple lines represent the contributions from fission yield, decay energy, decay constant and branching ratio respectively. The solid and dashed lines represent the results calculated with Set-A and Set-B.} 
    \label{fig-UQ-HLP-235U}
\end{figure}

\begin{figure}[!tb]
    \centering 
    \includegraphics[width=0.45\textwidth]{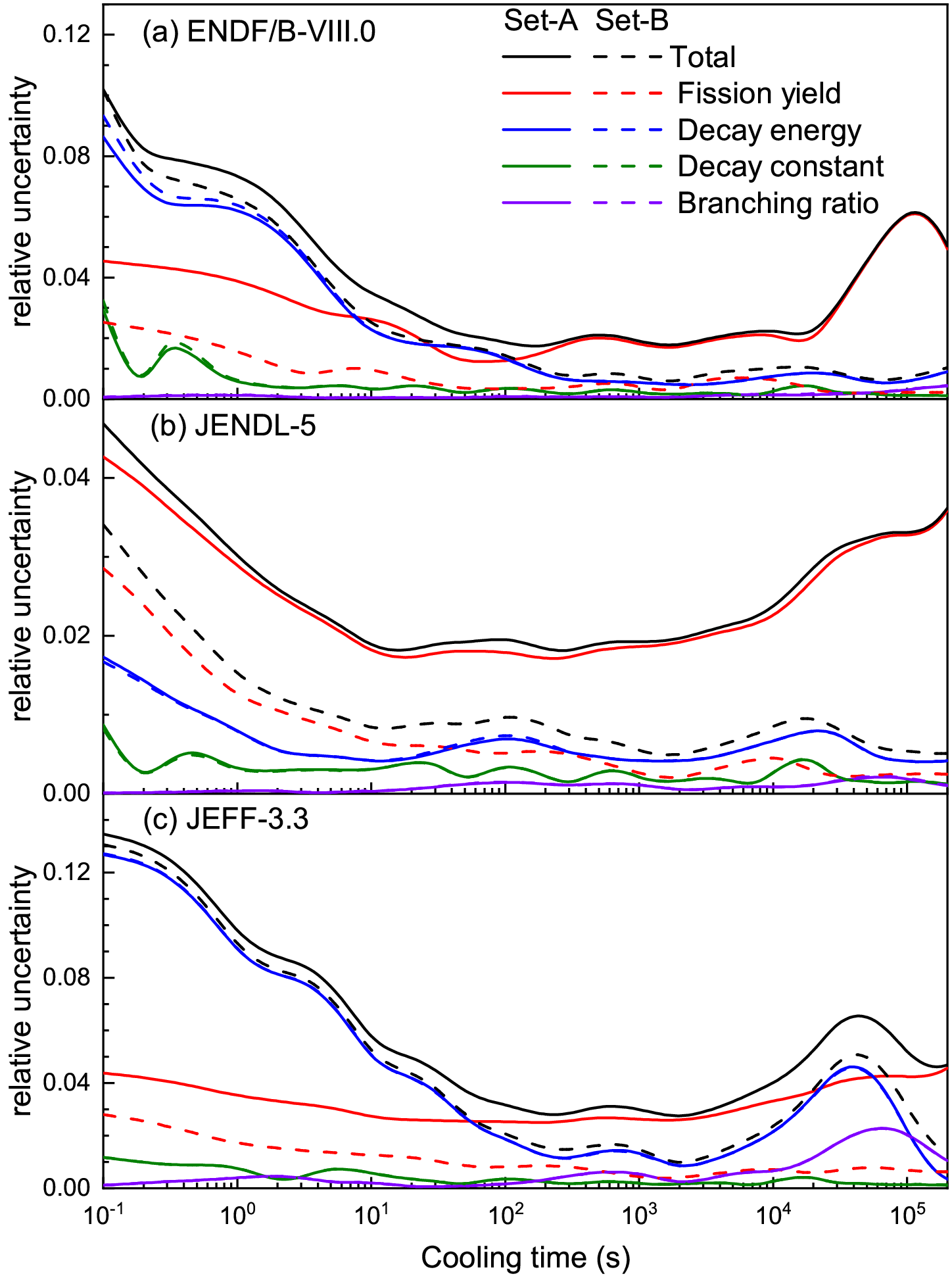}	
    \caption{(color online) Same as Fig. \ref{fig-UQ-HLP-235U} but for electromagnetic decay heat.} 
    \label{fig-UQ-HEM-235U}
\end{figure}

Fig. \ref{fig-UQ-HLP-235U} and \ref{fig-UQ-HEM-235U} show the relative uncertainties of light particle and electromagnetic decay heats for $^{235}$U($n_{\mathrm{th}},f$), including contributions of each kind of nuclear data. The uncorrelated yield data in Set-A provides a $\sim 4\%$ uncertainty in all cases, dominating the total uncertainty at cooling times longer than \SI{100}{\second}. On the contrary, the contribution from yield data is strongly reduced in Set-B and it decreases gradually as time evolves, which are resulted by the generated covariance matrix. 

The uncertainties propagated from decay energy, decay constant and branching ratio are almost unchanged in Set-A and Set-B results as decay data have not been adjusted. In general, decay energy is the primary contributor among the three kinds of decay data. It is noteworthy that the decay energy uncertainty obtained with JENDL-5 is much smaller than those obtained with ENDF/B-VIII.0 and JEFF-3.3 at cooling times shorter than \SI{100}{\second}. This is largely because that JENDL-5 provides decay energy uncertainties for more radionuclides than ENDF/B-VIII.0 and JEFF-3.3. These radionuclides are mainly short-lived and release energies in the early cooling times. Therefore the assumed 100$\%$ uncertainties are less adopted for JENDL-5 decay energy data, resulting in a smaller uncertainty.

With the generated covariance of yield data, the uncertainties of light particle and electromagnetic decay heat at cooling time \SI{0.1}{\second} are about 10$\%$ for ENDF/B-VIII.0 and JEFF-3.3 and that is 5$\%$ for JENDL-5. At cooling times longer than 10$^{5}$ s, the uncertainties decrease to about 1$\%$ for all three libraries. The main component of uncertainty for light particle decay heat comes from the decay energy uncertainty. This situation keeps for the electromagnetic decay heat uncertainty obtained with ENDF/B-VIII.0 and JEFF-3.3, while the uncertainties of yield data and decay energy dominate the total uncertainty of electromagnetic heat at cooling times shorter and longer than \SI{50}{\second} respectively for JENDL-5. 

\subsection{Contributions of important fission products and their sensitive coefficients}
\label{sec-3-3}

As GLS updating procedure generates covariance and adjusts IFYs simultaneously, we are concerned about its influence on the contributions of important fission products to decay heat as well as their sensitive coefficients. Analysis was performed at cooling time \SI{10}{\second}, which is the position of the curve peak for light particle heat as shown in Fig. \ref{fig-HLP-235U}. Moreover, an underestimation was given by JEFF-3.3 for electromagnetic heat around this time as presented in Fig. \ref{fig-HEM-235U}.

\begin{figure}[!tb]
    \centering 
    \includegraphics[width=0.45\textwidth]{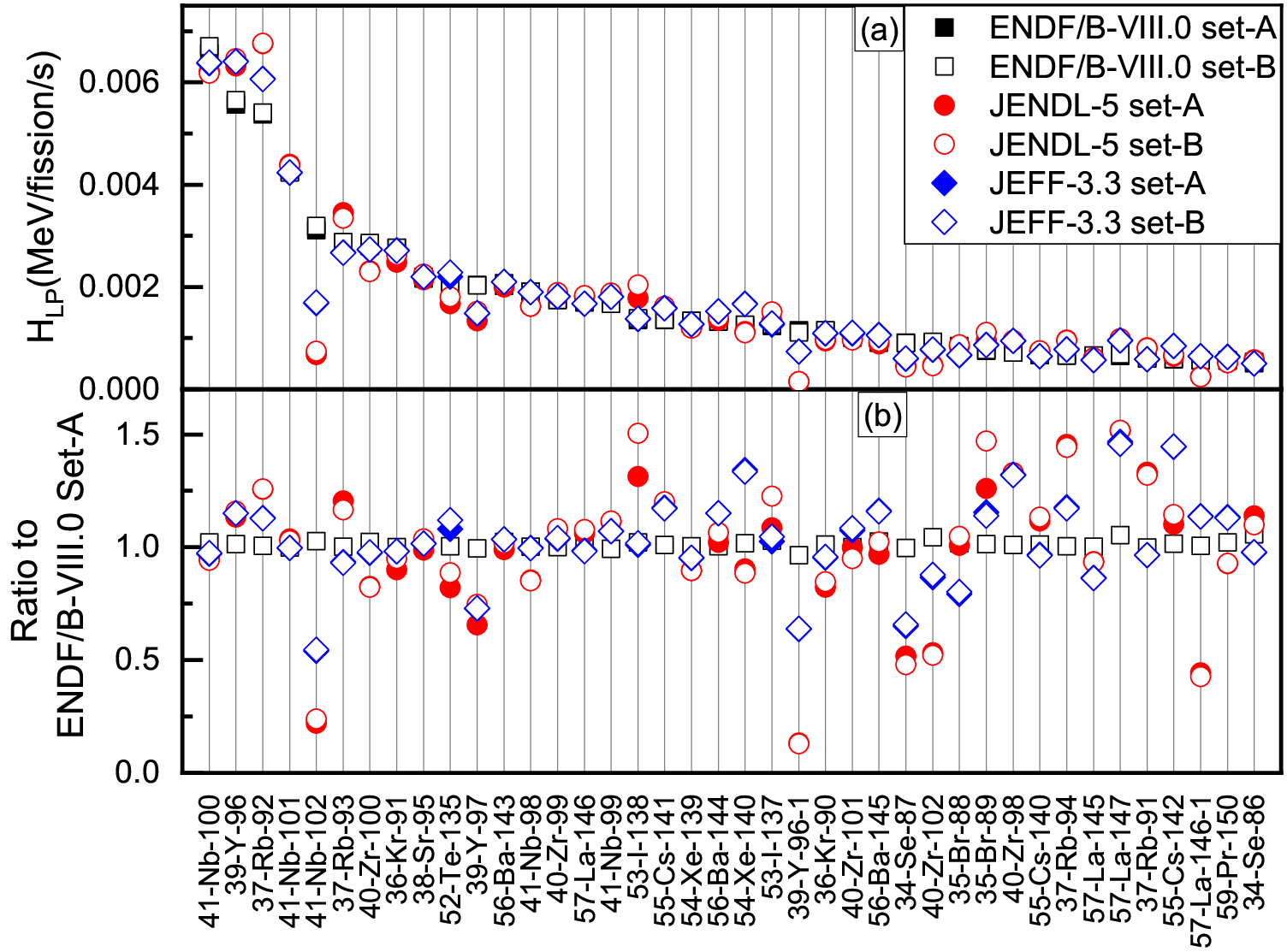}	
    \caption{(color online) Dominant contributors to light particle decay heat of $^{235}$U($n_{\mathrm{th}},f$) at cooling time \SI{10}{\second}. (a) Contributions of important fission products calculated with different sets of data. (b) Ratio of contribution of important fission product to that obtained with ENDF/B-VIII.0 Set-A. Nuclide is denoted in the form $Z$-[element name]-$A$ or $Z$-[element name]-$A$-$M$ if it is in ground state or isomeric state. See text for details. }
    \label{fig-HLP-component-235U}
\end{figure}

A comparison of 40 dominant contributors is shown in Fig. \ref{fig-HLP-component-235U} for light particle decay heat. These nuclides are sorted according to their contributions obtained with ENDF/B-VIII.0 Set-A. In all cases, over 80$\%$ light particle decay heat is provided by these 40 nuclides at cooling time \SI{10}{\second}. Different from the reasonable agreement in $H_{\mathrm{LP}}$, visible divergences occur among the contributions of important fission products obtained with different evaluated libraries. Especially, the contributions of $^{102}$Nb, $^{96m}$Y and $^{146m}$La calculated with JENDL-5 are less than half of those calculated with ENDF/B-VIII.0 and JEFF-3.3, no matter which set of data is used. In addition, the adjustment on IFYs has a weak influence on the contributions for ENDF/B-VIII.0 and JEFF-3.3, while the contributions of some specific nuclides are affected clearly for JENDL-5, such as $^{97}$Y, $^{135}$Te, $^{137,138}$I and $^{89}$Br.

\begin{figure}[!tb]
    \centering 
    \includegraphics[width=0.45\textwidth]{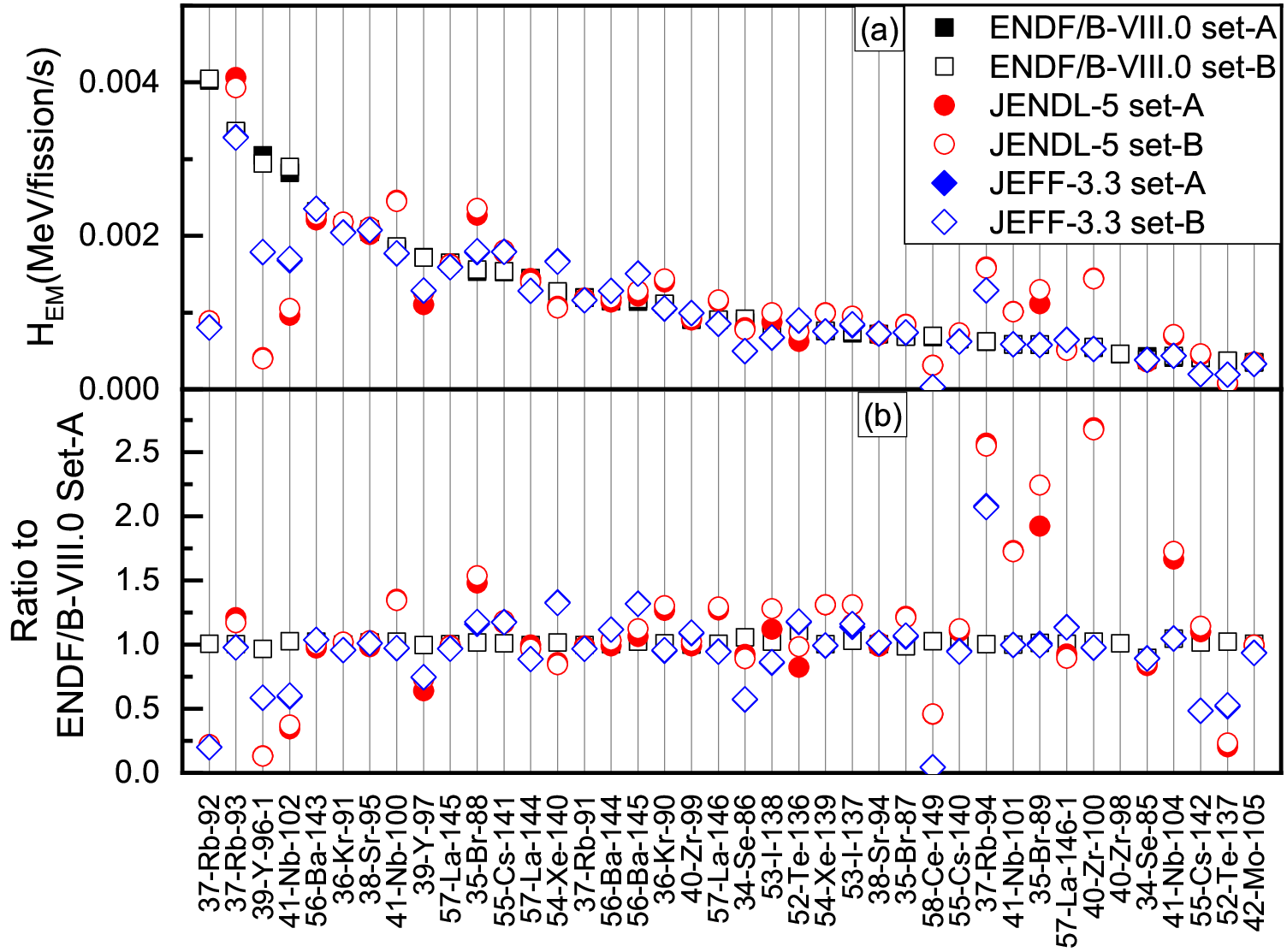}	
    \caption{(color online) Same as Fig. \ref{fig-HLP-component-235U} but for electromagnetic decay heat. }
    \label{fig-HEM-component-235U}
\end{figure}

More significant divergences can be observed for electromagnetic decay heat as presented in Fig. \ref{fig-HEM-component-235U}. These 40 nuclides are the most significant contributors in the calculations with ENDF/B-VIII.0 Set-A and their total contribution accounts for over 75$\%$ electromagnetic decay heat in all cases. A noteworthy thing is that $^{98}$Zr has no contribution to electromagnetic heat if the decay data of JENDL-5 or JEFF-3.3 are used, while it contributes 0.8$\%$ in the calculations with ENDF/B-VIII.0. It is because that $\bar{E}_{\mathrm{EM}}$ of $^{98}$Zr is set to be zero in JENDL-5 and JEFF-3.3 while that is \SI{0.449}{\MeV} in ENDF/B-VIII.0. The contributions calculated with JEFF-3.3 are lower than those obtained with other two libraries for many nuclides, resulting in the underestimation of electromagnetic heat around cooling time \SI{10}{\second}. Comparing the results obtained with Set-A and Set-B, the influence of IFYs adjustment is only considerable for JENDL-5, such as $^{138}$I, $^{136}$Te and $^{89}$Br.

Relative sensitivity coefficient of decay heat ($H$) to nuclear data ($x$) is defined as
\begin{equation}\label{e-rela-sensi}
s_H\left( x \right) =\frac{\partial H/H}{\partial x/x}, 
\end{equation}
which is slightly different from $S_H$ given in Eq. (\ref{e-sensi}). It has been discussed in Sec. \ref{sec-3-2} that the decay heat uncertainties propagated from decay data are almost unchanged in Set-A and Set-B calculations. Therefore we here focus on the relative sensitivity coefficient to IFYs, that is $s_H(Y_{\mathrm{I}})$. 

\begin{table}[!tb]
\centering
\footnotesize
\begin{tabular}{l c c c c c c} 
 \hline
  Nuclide & \multicolumn{2}{c}{ENDF/B-VIII.0} & \multicolumn{2}{c}{JENDL-5} & \multicolumn{2}{c}{JEFF-3.3} \\         
                   & Set-A & Set-B & Set-A & set-B & Set-A & Set-B  \\
 \hline
$^{96}$Y   &     - &    - & 2.02 & 1.73 &    - &    - \\
$^{92}$Rb  &  3.64 & 3.59 & 4.81 & 4.41 & 3.58 & 3.57 \\
$^{101}$Nb &  1.70 & 1.67 & 1.78 & 1.89 & 1.37 & 1.34 \\
$^{93}$Rb  &  2.98 & 2.94 & 3.88 & 3.64 & 2.50 & 2.48 \\
$^{100}$Zr & 10.13 & 9.78 & 9.40 & 9.06 & 9.01 & 9.01 \\
$^{91}$Kr  &  3.48 & 3.43 & 3.46 & 3.58 & 3.36 & 3.35 \\
$^{95}$Sr  &  2.31 & 2.32 & 2.31 & 2.38 & 2.25 & 2.24 \\
$^{135}$Te &  2.41 & 2.39 & 2.02 & 2.15 & 2.52 & 2.62 \\
$^{97}$Y   &  1.21 & 1.17 &    - &    - &    - &    - \\
$^{143}$Ba &  1.81 & 1.82 & 1.89 & 1.84 & 1.70 & 1.70 \\
$^{99}$Zr  &  2.27 & 2.20 & 2.45 & 2.27 & 2.27 & 2.25 \\
$^{146}$La &     - &    - & 1.25 & 1.18 &    - &    - \\
$^{138}$I  &  1.59 & 1.59 & 2.11 & 2.38 & 1.48 & 1.49 \\
$^{141}$Cs &  1.15 & 1.15 & 1.36 & 1.39 & 1.11 & 1.10 \\
$^{139}$Xe &  1.43 & 1.41 & 1.32 & 1.26 & 1.32 & 1.32 \\
$^{144}$Ba &  1.81 & 1.79 & 1.85 & 1.85 & 1.85 & 1.85 \\
$^{140}$Xe &  1.80 & 1.80 & 1.57 & 1.49 & 2.24 & 2.23 \\
$^{137}$I  &  1.36 & 1.38 & 1.58 & 1.74 & 1.31 & 1.34 \\
$^{96m}$Y  &  1.42 & 1.35 &    - &    - &    - &    - \\
$^{90}$Kr  &  1.36 & 1.35 & 1.17 & 1.16 & 1.26 & 1.26 \\
$^{101}$Zr &  4.29 & 4.23 & 4.62 & 4.27 & 4.48 & 4.51 \\
$^{145}$Ba &  1.45 & 1.47 & 1.44 & 1.49 & 1.57 & 1.57 \\
$^{87}$Se  &  1.12 & 1.09 &    - &    - &    - &    - \\
$^{102}$Zr &  3.65 & 3.61 & 2.76 & 2.64 & 2.78 & 2.81 \\
$^{89}$Br  &     - &    - & 1.21 & 1.39 &    - &    - \\
$^{98}$Zr  &  1.48 & 1.28 & 1.39 & 1.20 & 1.65 & 1.63 \\
$^{94}$Rb  &     - &    - & 1.27 & 1.23 &    - &    - \\
$^{147}$La &     - &    - & 1.11 & 1.07 &    - &    - \\
 \hline 
\end{tabular}
\caption{Relative sensitivity coefficients of light particle decay heat to independent fission yield for important fission products of $^{235}$U($n_{\mathrm{th}},f$) at cooling time 10 s. Coefficients are not presented if their values are both less than 0.01 in the analysis with Set-A and Set-B, and the nuclide is omitted if its $s_H(Y_{\mathrm{I}})$ is smaller than 0.01 in all cases. The rest are multiplied by 100 for ease of presentation. See text for details.}
\label{table-rela-sensi-HLP-yield}
\end{table}

\begin{table}[!tb]
\centering
\footnotesize
\begin{tabular}{l c c c c c c} 
 \hline
  Nuclide & \multicolumn{2}{c}{ENDF/B-VIII.0} & \multicolumn{2}{c}{JENDL-5} & \multicolumn{2}{c}{JEFF-3.3} \\         
                   & Set-A & Set-B & Set-A & set-B & Set-A & Set-B  \\
 \hline
$^{92}$Rb  & 3.77 & 3.72 &    - &    - &    - &    - \\
$^{93}$Rb  & 4.90 & 4.85 & 6.16 & 4.79 & 6.16 & 5.78 \\
$^{96m}$Y  & 5.19 & 4.95 & 0.68 & 3.28 &    - &    - \\
$^{102}$Nb &    - &    - & 1.60 & 0.10 & 1.60 & 1.71 \\
$^{143}$Ba & 2.78 & 2.80 & 2.76 & 2.93 & 2.76 & 2.69 \\
$^{91}$Kr  & 4.27 & 4.22 & 4.22 & 4.48 & 4.22 & 4.36 \\
$^{95}$Sr  & 3.02 & 3.04 & 2.89 & 3.25 & 2.89 & 2.97 \\
$^{97}$Y   & 1.41 & 1.37 &    - &    - &    - &    - \\
$^{145}$La & 1.50 & 1.46 & 1.36 & 1.07 & 1.36 & 1.25 \\
$^{88}$Br  & 2.06 & 2.04 & 3.28 & 2.61 & 3.28 & 3.38 \\
$^{141}$Cs & 1.79 & 1.79 & 2.02 & 1.92 & 2.02 & 2.07 \\
$^{144}$La & 2.50 & 2.50 & 2.01 & 3.42 & 2.01 & 1.91 \\
$^{91}$Rb  & 1.08 & 1.06 & 1.07 & 1.13 & 1.07 & 0.97 \\
$^{144}$Ba & 3.25 & 3.22 & 3.13 & 3.61 & 3.13 & 3.14 \\
$^{145}$Ba & 3.16 & 3.21 & 3.27 & 4.30 & 3.27 & 3.39 \\
$^{90}$Kr  & 1.85 & 1.84 & 2.29 & 1.92 & 2.29 & 2.27 \\
$^{99}$Zr  & 1.06 & 1.03 & 1.06 & 1.23 & 1.06 & 0.98 \\
$^{146}$La &    - &    - & 1.06 & 0.37 & 1.06 & 1.00 \\
$^{86}$Se  & 1.18 & 1.17 & 1.65 & 1.15 & 1.65 & 1.56 \\
$^{138}$I  & 1.27 & 1.27 & 1.38 & 1.13 & 1.38 & 1.56 \\
$^{136}$Te & 1.35 & 1.46 & 1.15 & 1.83 & 1.15 & 1.35 \\
$^{139}$Xe & 1.11 & 1.09 & 1.45 & 1.20 & 1.45 & 1.38 \\
$^{137}$I  & 1.08 & 1.10 & 1.30 & 1.32 & 1.30 & 1.43 \\
$^{94}$Sr  &    - &    - & 0.83 & 1.02 &    - &    - \\
$^{87}$Br  &    - &    - & 1.07 & 1.02 & 1.07 & 1.06 \\
$^{149}$Ce & 1.07 & 1.07 &    - &    - &    - &    - \\
$^{94}$Rb  & 1.28 & 1.27 & 2.80 & 2.35 & 2.80 & 2.72 \\
$^{89}$Br  & 1.02 & 1.02 & 1.92 & 1.05 & 1.92 & 2.21 \\
$^{146m}$La &    - &    - & 0.87 & 1.18 &    - &    - \\
$^{100}$Zr & 3.53 & 3.41 & 5.74 & 3.51 & 5.74 & 5.53 \\
 \hline 
\end{tabular}
\caption{Same as Table \ref{table-rela-sensi-HLP-yield} but for electromagnetic decay heat. See text for details.}
\label{table-rela-sensi-HEM-yield}
\end{table}

Compilations of $s_H(Y_{\mathrm{I}})$ are listed in Table \ref{table-rela-sensi-HLP-yield} and \ref{table-rela-sensi-HEM-yield} for the important fission products which have been presented in Figs. \ref{fig-HLP-component-235U} and \ref{fig-HEM-component-235U} respectively. The nuclide is omitted if its $s_H(Y_{\mathrm{I}})$ is smaller than 0.01 in all calculations. Generally, $^{100}$Zr and $^{93}$Rb have the highest relative sensitivity coefficients for light particle and electromagnetic decay heats respectively among the three evaluated libraries. The relative changes due to the yield adjustment are smaller than 20$\%$ for $s_H(Y_{\mathrm{I}})$ of two kinds of decay heats in most cases, revealing a weak influence. However dramatic changes are observed for the relative sensitivity coefficients of electromagnetic decay heat calculated with JENDL-5. For example, the coefficient of $^{96m}$Y is enhanced by about 5 times after GLS updating process and those of $^{102}$Nb and $^{146}$La are reduced by about 95$\%$ and 65$\%$ respectively.

\section{Summary and conclusions}
\label{sec-4}

The generalized least squares updating approach was implemented in covariance generation for $^{235}$U($n_{\mathrm{th}},f$) independent fission yields in the widely used evaluated nuclear data libraries ENDF/B-VIII.0, JENDL-5 and JEFF-3.3, with the constraints of basic physical conservation equation and chain yield data. The variances of original IFYs were strongly reduced and they were reintroduced as correlations between IFYs, which are mainly negative correlations. The yield data were also slightly adjusted in GLS updating process simultaneously.

The original and updated IFYs as well as decay data were used in the uncertainty analyses of decay heat by means of generalized perturbation theory, including the uncertainties propagated from yield and decay data. Good agreement on decay heat was achieved by three libraries at cooling time longer than \SI{1000}{\second}. However, discrepancies were found within \SI{1000}{\second} and a significant underestimation was resulted by JEFF-3.3 for electromagnetic decay heat from 5 to 50 \si{\second}. The adjustment of IFYs has a weak influence on decay heat, while the generated correlation strongly reduces the contribution of yield data to decay heat uncertainty. Using the updated yield data with covariance matrixes, the uncertainties of decay heat are $\sim$10$\%$ for ENDF/V-VIII.0 and JEFF-3.3 and $\sim$5$\%$ for JENDL-5 at cooling time \SI{0.1}{\second} and those are about 1$\%$ for three libraries at \SI{1E5}{\second}, mainly coming from decay energy data. The influence of yield data adjustment on decay heat contributions of important fission products as well as their sensitive coefficients has also been investigated. It is found that the influence is little for ENDF/B-VIII.0 and JEFF-3.3 while noticeable changes occur on some important fission products for JENDL-5. 

The validness of the present fission yield covariance generation approach was justified and the updated fission yield data are expected to be applicable in nuclear design and safety analysis together with their covariance matrixes. 

\appendix
\section{Covariance matrix for branching ratios}
\label{App-1}

For a set of gaussian random variables $ \left\{ x_1,x_2,...,x_n \right\} $ with mean values $\left\{ \mu _1,\mu _2,...,\mu _n \right\} $ and variances $\left\{ \sigma _{1}^{2},\sigma _{2}^{2},...,\sigma _{n}^{2} \right\} $, their joint distribution $f\left( x_1,x_2,...,x_n \right)$ is given as following if there is a summation constraint $x_1+x_2+...+x_n=x$, 
\begin{equation}\label{e-App-joint-distribution}
\begin{aligned}
f = & N\int{dx}\delta \left( x_1+x_2+...+x_n-x \right) 
\\
& \times \exp \left[ -\frac{\left( x-\mu \right) ^2}{2\sigma ^2} \right] \prod_{i=1}^n{\exp \left[ -\frac{\left( x_i-\mu _i \right) ^2}{2\sigma _{i}^{2}} \right]},
\end{aligned}
\end{equation} 
where $N$ is the normalization factor. The sum $x$ is also a gaussian random variable with mean value $\mu$ and variance $\sigma ^2$.

Due to the summation constraint, the variance of $x_i$ should be adjusted as
\begin{equation}\label{e-App-diag-adjust}
\begin{aligned}
\tilde{\sigma}_{i}^{2} = & E\left( x_{i}^{2} \right) -E^2\left( x_i \right) 
\\
= & \int{dx_1dx_2...dx_n}x_{i}^{2}f-\left[ \int{dx_1dx_2...dx_n}x_if \right] ^2
\\
= & \sigma _{i}^{2}\left[ 1-\frac{\sigma _{i}^{2}}{\sigma ^2+r} \right], \quad r=\sum_{i=1}^n{\sigma _{i}^{2}}, 
\end{aligned}
\end{equation} 
corresponding to the diagonal element of covariance matrix $\boldsymbol{V}_{ii}$. The non-diagonal element $\boldsymbol{V}_{ij}$ is given as
\begin{equation}\label{e-App-non-diag-adjust}
\begin{aligned}
\boldsymbol{V}_{ij} = & E\left( x_ix_j \right) -E\left( x_i \right) E\left( x_j \right) 
\\
= & \int{dx_1dx_2...dx_n}x_ix_jf
\\
& -\int{dx_1dx_2...dx_n}x_if\times \int{dx_1dx_2...dx_n}x_jf
\\
= & -\frac{\sigma _{i}^{2}\sigma _{j}^{2}}{\sigma ^2+r}, i \neq j.
\end{aligned}
\end{equation} 

As the branching ratios strictly satisfy the normalization condition $\sum_j{b_{ji}}=1$, Eq. (\ref{e-App-diag-adjust}) and (\ref{e-App-non-diag-adjust}) should be used to generate their covariance matrix with $\sigma$=0 as shown in Eq. (\ref{e-cov-BR}), which means that the gaussian distribution of $x$ approaches its limit and becomes a dirac delta distribution.

\end{document}